\documentclass[aps,prb,twocolumn]{revtex4} 

\usepackage{graphicx}
\usepackage{dcolumn}
\usepackage{bm}
\usepackage{amsmath}  

\newcommand{\comment}[1]{}

\begin{document}
\renewcommand{\theequation}{\arabic{section}.\arabic{equation}}

\title{On the Fountain Effect in Superfluid Helium}


\author{Phil Attard}
\affiliation{
{\tt phil.attard1@gmail.com}}
\noindent {\tt  Projects/QSM22/Paper3/fountain.tex}


\begin{abstract}
The origin of the fountain pressure
that arises from a small temperature difference in superfluid helium is analyzed.
The osmotic pressure explanation due to Tisza (1938),
based on the different fractions of ground state bosons,
is formulated as an ideal solution
and shown to overestimate the fountain pressure by an order of magnitude or more.
The experimentally confirmed thermodynamic expression of H. London (1939),
in which the temperature derivative of the pressure
equals the entropy per unit volume,
is shown to be equivalent to equality of chemical potential,
not chemical potential divided by temperature.
The former results from minimizing the energy at constant entropy,
whereas the latter results from maximizing the entropy.
It is concluded that superfluid flow in the fountain effect
is driven by mechanical, not statistical, forces.
This new principle appears to be a general law
for flow in superfluids and superconductors.
\end{abstract}


\maketitle

%
\section{Introduction}
\setcounter{equation}{0} \setcounter{subsubsection}{0}
%

One of the earliest and perhaps still the most spectacular
manifestation of superfluidity is the fountain effect.
In this helium vigorously spurts from the open end
of a heated  tube
that is connected by a capillary or microporous frit
to a chamber of liquid helium
maintained below the condensation temperature.\cite{Allen38,Balibar17}
Closing the heated chamber
and measuring the fountain pressure
is a common experimental technique for obtaining the entropy.\cite{Donnelly98}

The commonly accepted physical explanation for the fountain effect
as an osmotic pressure is due to Tisza.\cite{Balibar17,Tisza38}
It is based on his two-fluid model,
which in turn is based on F. London's proposal
that the $\lambda$-transition in liquid helium
is due to Bose-Einstein condensation.\cite{FLondon38}
The two-fluid model says that helium consists of a mixture of He I and He II,
which are excited state and ground state bosons, respectively.
The latter comprise the superfluid
and their fraction increases with decreasing temperature
below the condensation temperature.
This gives rise to the notion
that osmosis drives He II selectively through the capillary 
from the low temperature, high concentration chamber
to the high temperature, low concentration one.
In general in a binary solution the mixing entropy favors concentration equality,
and this is what is said to create the osmotic pressure.

H. London\cite{HLondon39} carried out a thermodynamic analysis
of the fountain effect that made a quantitative prediction
for the pressure difference for a given temperature difference.
His result is historically important
as the quantitative experimental verification of this formula
was evidence for the picture of superfluid helium
being in a state of zero entropy.
To the present day his expression remains significant
as fountain pressure measurements
are used to establish benchmark results
against which calorimetric methods for the entropy of helium
may be tested.\cite{Donnelly98}
The results obtained from his expression are also used
as calibration standards for instruments.\cite{Hammel61}

Although this paper  does not go beyond
the expression of H. London,\cite{HLondon39}
it does shed significant new light
on the physical origin of the fountain effect
and on the nature of superfluidity.
The major result is to show that the H. London expression
is obtained by minimizing the energy of the system
with respect to particle transfer between the two chambers,
Sec.~\ref{Sec:minE}.
Conversely, it shows that the result
is inconsistent with maximizing the entropy of the system.
This apparent violation of the Second Law of Thermodynamics
is peculiar to superfluid flow.

The new  derivation and interpretation
of the H. London expression given here
has a number of consequences for superfluid flow.
One is that it rules out the osmotic pressure explanation of the fountain effect
due to Tisza.\cite{Tisza38}
The latter is statistical and based on the mixing entropy,
which contradicts the present finding that it is
the minimization of energy that underlies the fountain effect.
To underscore this point,
in Sec.~\ref{Sec:Osmo}
quantitative calculations are made
of the osmotic prediction for the fountain pressure
using both calculated\cite{Attard21,Attard22a}
and measured\cite{Donnelly98} data for the condensed boson fraction.
It is shown that the predicted fountain pressures
do not agree with measured values.\cite{Hammel61}
Although these calculations are made at the level of ideal solution theory,
the quantitative error is so large,
and the qualitative difference with energy minimization so stark,
that the osmotic explanation for the fountain effect appears nonviable.

This paper is the fourth in a series
on superfluidity and superconductivity.
\cite{Attard22a,Attard22b,Attard22c}
The general formulation of quantum statistical mechanics
that underlies them is given in Ref.~\onlinecite{Attard21}.
The author has also given coherent formulations
of equilibrium\cite{TDSM}
and non-equilibrium\cite{NETDSM}
thermodynamics and classical statistical mechanics,
which results and concepts are freely used below.

%
\section{Thermodynamic Analysis} \label{Sec:TD}
\setcounter{equation}{0} \setcounter{subsubsection}{0}
%

\subsection{The Fountain Pressure Minimizes Energy} \label{Sec:minE}

The aim of this section is show that
the expression of H. London\cite{HLondon39} for the fountain pressure
follows from the minimization of the energy of the subsystems.

Consider two closed chambers of helium,
$A$ and $B$,
each in contact with its own thermal reservoir of temperature
$T_A$ and $T_B$,
and having pressure $p_A$ and $p_B$.
The chambers are connected by a capillary through which superfluid,
and only superfluid, flows.
Chamber $A$ in practice is at the lower temperature,
and consists of saturated liquid and vapor,
but these points are presently unimportant.
As H. London points out,\cite{HLondon39}
in the optimum steady state
the pressure of the second chamber is a function of its temperature
and the pressure and temperature of the first chamber,
$p_B = p(T_B;p_A,T_A)$.

The result given by H. London\cite{HLondon39}
says that the derivative of the pressure of the second chamber
with respect to its temperature for fixed first chamber
equals the entropy density,\cite{HLondon39}
\begin{equation} \label{Eq:HLondon-dp/dT}
\frac{\mathrm{d}p_B}{\mathrm{d}T_B} = \rho_B s_B .
\end{equation}
Here $\rho$ is the number density and $s$ is the entropy per particle.
(In this paper I use lower case letters to denote quantities per particle;
H. London\cite{HLondon39} uses them to denote quantities per unit mass.)
(The derivative of the pressure and like quantities
with respect to temperature for fixed first chamber parameters
are total derivatives,
which means that their integral along the fountain path
can be evaluated analytically.
This is useful in the analysis of the H. London derivation\cite{HLondon39}
in Appendix~\ref{Sec:antiLondon}.)

Whereas H. London\cite{HLondon39} purported to derive this result
using a work-heat flow cycle (see Appendix~\ref{Sec:antiLondon}),
here I seek to locate it within the broader principles of  thermodynamics.
To this end there is a subtle point regarding extensivity
that needs to be clarified.

In equilibrium thermodynamics it is generally recognized
that quantities such as energy,  entropy, and  free energy are extensive,
which means that the total value is the sum of the individual values
for independent or quasi-independent subsystems.
For the present case of two subsystems of different temperatures
this general rule does not hold for the free energy.
This is easily seen by considering the two to be quasi-independent
such that the probability of a joint state is the product of the
individual probabilities\cite{TDSM}
\begin{eqnarray}
\wp(X_A,X_B) & = &
\wp_A(X_A) \wp_B(X_B)
\nonumber \\ & \propto &
e^{ S_A(X_A)/k_\mathrm{B} } e^{ S_B(X_B)/k_\mathrm{B} }
\nonumber \\ & = &
e^{ - F_A(X_A)/k_\mathrm{B}T_A } e^{ -F_B(X_B)/k_\mathrm{B}T_B } .
\end{eqnarray}
Here $k_\mathrm{B}$ is Boltzmann's constant,
and $S$ and $F$ are the appropriate total entropy and free energy.
From this it is clear that the entropy is simply additive,
$S_\mathrm{tot}(X_A,X_B) = S_A(X_A) + S_B(X_B) $,
but it is the free energy divided by temperature,
rather than the free energy itself, that is additive,
$F_\mathrm{tot}(X_A,X_B)/T_\mathrm{eff} = F_A(X_A)/T_A + F_B(X_B)/T_B $.
It is only in the case of subsystems all with the same temperature
that the free energy is extensive (i.e.\ simply additive).
By the usual principles of mechanics,
the energy is simply additive
$E_\mathrm{tot}(X_A,X_B) = E_A(X_A) + E_B(X_B) $,
which result will prove essential for the following analysis.

In searching for the thermodynamic basis
of the H. London expression,\cite{HLondon39} Eq.~(\ref{Eq:HLondon-dp/dT}),
there are two possible axioms to be considered.
The first is that the total entropy of the total system
is a maximum, which is of course just the Second Law of Thermodynamics,
albeit applied to a non-equilibrium steady state system.\cite{NETDSM}
The second is that the total energy of the subsystems be minimized,
which is a mechanical law that is generally without thermodynamic relevance,
the present case being the only significant exception
that I can think of.

A third possible axiom, that the total free energy of the total system
is a minimum is not viable for two related reasons.
First, as just explained,
it is not free energy but free energy divided by temperature
that is the relevant thermodynamic potential,
so it should be the sum of the free energy divided by temperature
that is a minimum.
And second, the free energy  divided by temperature
is derived directly from the total entropy,
and so the principle of free energy divided by temperature minimization
is not separate to, and does not yield any new information
beyond, total entropy maximization.
This can be seen by the trivial relationship
between the second and third equalities in the above equation.
To be clear and unambiguous on this point:
for the case of systems with different temperatures,
there is no principle of free energy minimization.
There is a principle of  free energy divided by temperature minimization,
but this is no different to the principle of entropy maximization.

Consider now the first possible axiom,
that the total entropy of the system is a maximum.
Since the systems are closed, the total entropy is\cite{TDSM}
\begin{equation}
S_\mathrm{tot} =
\frac{- F(N_A,V_A,T_A)}{T_A} -  \frac{F(N_B,V_B,T_B)}{T_B} ,
\end{equation}
where $N$ is the number, $V$ is the volume,
and $F$ is the Helmholtz free energy.
(The total subsystem-dependent entropy
of a sub-system and its reservoir is the negative of the free energy
divided by the temperature.)\cite{TDSM}
With the total number of helium atoms fixed, $N=N_A+N_B$,
its derivative is\cite{TDSM}
\begin{equation}
\frac{\partial S_\mathrm{tot}}{\partial N_A}
 =
\frac{-\mu_A}{T_A} +  \frac{\mu_B}{T_B} ,
\end{equation}
where $\mu$ is the chemical potential.
The maximum total entropy occurs when this is zero,
which gives the condition for the optimum steady state
that follows from the first possible axiom as
\begin{equation} \label{Eq:mu/T}
\frac{\mu_A}{T_A} =  \frac{\mu_B}{T_B} .
\end{equation}

Consider now the second possible axiom,
that the total energy is a minimum.
The energy of each chamber is a function
of its entropy, volume, and number,
$E_\mathrm{tot} = E(S_A,V_A,N_A) +  E(S_B,V_B,N_B)$,
which is standard.\cite{TDSM}
In this case the derivative at fixed $N$ is\cite{TDSM}
\begin{equation}
\frac{\partial E_\mathrm{tot}}{\partial N_A}
 =
\mu_A  - \mu_B .
\end{equation}
Obviously the minimum energy state corresponds to
\begin{equation} \label{Eq:muA=muB}
\mu_A =  \mu_B .
\end{equation}
The principle of energy minimization dates to Isaac Newton (1687).
It applies to mechanical systems,
in which statistical or entropic considerations are absent.
Although one could obtain
this same result by minimizing the simple sum
of the free energies of the two chambers,
as discussed above such procedure is not correct
as it is actually the free energy divided by temperature
that is additive.

The chemical potential is the Gibbs free energy per particle,
$\mu = G(N,p,T)/N$.\cite{TDSM}
The derivative of Eq.~(\ref{Eq:muA=muB}) with respect to $T_B$
at constant pressure and temperature of the first chamber,
and number of the second, is\cite{TDSM}
\begin{eqnarray}
0
& = &
\frac{\mathrm{d}(G_B/N_B)}{\mathrm{d}T_B}
\nonumber \\ & = &
\frac{\partial g_B}{\partial T_B}
+
\frac{\partial g_B}{\partial p_B}
\frac{\mathrm{d}p_B}{\mathrm{d}T_B}
\nonumber \\ & = &
 -s_B + v_B \frac{\mathrm{d}p_B}{\mathrm{d}T_B} ,
\end{eqnarray}
where $g$, $s$, and $v = \rho^{-1}$
are the Gibbs free energy, entropy, and volume per particle,
respectively.
This is the same as H. London's expression,
Eq.~(\ref{Eq:HLondon-dp/dT}).
Conversely,
the derivative of the expression
that results from maximizing the total entropy, Eq.~(\ref{Eq:mu/T}),
does not yield this result.

Equation~(\ref{Eq:muA=muB}) is the major result of this paper:
on  the fountain path the  chemical potential is constant,
which is the same as saying that the energy
is minimized at constant entropy.

\subsubsection{Rationalization}

A possible way to understand the equality of chemical potential
in the fountain effect is as follows.
Each chamber consists of a mixture of ground state bosons,
denoted by subscript zero,
and excited state bosons,
denoted by subscript asterisk.
These are in equilibrium with each other,
and so we can drop the subscript on the chemical potential,
$\mu_{0A} = \mu_{*A} = \mu_A$,
and similarly for chamber $B$.
This follows because each chamber has uniform temperature.

We imagine that the transfer of $N$ condensed bosons
via the capillary from chamber $A$ to $B$
is accomplished by a contiguous packet
of initial volume $N v_{0A}$.
This packet has initial pressure, $p_A$, initial enthalpy $h_{0A}$,
and the condensed bosons within it have initial chemical potential
$\mu_{0A}$.
(The energy per particle and the volume per particle
are well-defined in a mixture and can be calculated by
statistical mechanical techniques.)
We suppose that the usual relationships of equilibrium thermodynamics
hold for the packet,
including that the chemical potential
is the Gibbs free energy per particle,\cite{TDSM}
\begin{equation}
\mu_{0A} = h_{0A} - T_{0A} s_{0A} ,
\end{equation}
where $s_{0A}$ is the initial entropy per boson in the packet.
A similar expression holds for $\mu_{0B}$.
It is not necessarily assumed that the initial or final temperatures
of the packet equal that of their chamber.

We assume that at the entrance and exit of the capillary
the bosons in the packet are in equilibrium with the bosons
in the respective chambers,
\begin{equation}
\mu_{0A} = \mu_A , \mbox{ and } \mu_{0B} = \mu_B.
\end{equation}

We assume that no heat flows into the packet during its transit so that
the enthalpy is constant,
\begin{equation}
h_{0A} = h_{0B} .
\end{equation}
(The absence of heat flow is the reason
that one cannot assume temperature equality at the termini.)

From these three assumptions it follows that
\begin{equation}
\mu_A  - \mu_B
=
- T_{0A} s_{0A}  + T_{0B} s_{0B}  .
\end{equation}
If one takes the superfluid condition to be that
the entropy of the condensed bosons is zero,
$s_{0A} = s_{0B} = 0$,
then this says that the chemical potentials of the two chambers
are equal, $\mu_A  = \mu_B$.

Conversely, if  the chemical potentials  are equal,
then the products on the right hand side
must be a constant on the fountain path,
\begin{equation}
T_{0A} s_{0A}  = T_{0B} s_{0B}  = f(\mu_A)  .
\end{equation}
If the internal temperature of the packet
is a non-constant function of the chamber temperature, $T_{0B}(T_B)$,
then because $f(\mu_A)$ does not vary with the second chamber temperature,
one must have that
\begin{equation}
s_{0B} = f(\mu_A) = 0.
\end{equation}

Since the pressure of the packet at its terminus is $p_B$,
which varies with $T_B$,
it is difficult to see how the internal temperature
could be independent of the chamber temperature.
Realistically,
the only possible value for a constant internal temperature would be zero.
But this cannot be its value
because the enthalpy goes to zero at absolute zero, $h_{0B}(T_{0B}=0)=0$,
at least according to the generally accepted
Nernst heat theorem\cite{HLondon39}
and measured saturation\cite{Donnelly98} data.
The vanishing of the enthalpy and entropy
would mean that the chemical potential also vanished,
$  \mu _{0B} = h_{0B} + T_{0B} s_{0B} = 0$,
whereas the equality of chemical potentials means that
in general it is non-zero,
$  \mu _{0B} = \mu_{B} = \mu _{A} \ne 0$.

\subsection{Numerical  and Experimental Results}

In practice in experimental application,\cite{Hammel61}
the H. London expression for the derivative of the fountain pressure,
Eq.~(\ref{Eq:HLondon-dp/dT}),
is integrated along the saturation curve,
\begin{equation} \label{Eq:HL-int}
p_B - p_A =
\int_{T_A}^{T_B} \mathrm{d}T'\;
\rho^\mathrm{sat}(T') s^\mathrm{sat}(T')   .
\end{equation}

Since the integral for the fountain pressure should be evaluated
on the fountain path rather than the saturation path,
one can correct this as follows.
One has, with pressure and temperature as the independent variables,
\begin{eqnarray}
\rho s & = &
\rho^\mathrm{sat}  s^\mathrm{sat}
+ (p-p^\mathrm{sat})
\left(\frac{\partial (\rho s) }{\partial p} \right)_T^\mathrm{sat}
\nonumber \\ & \approx &
\rho^\mathrm{sat}  s^\mathrm{sat}
+ (p-p^\mathrm{sat}) \rho^\mathrm{sat}
\left(\frac{1}{T} \frac{\partial h }{\partial p}
- \frac{v}{T}  \right)_T^\mathrm{sat}
\nonumber \\ & = &
\rho^\mathrm{sat}  s^\mathrm{sat}
- \alpha^\mathrm{sat} (p-p^\mathrm{sat}) .
\end{eqnarray}
In the second equality the liquid has been taken to be incompressible.
Here the thermal expansivity is
$ \alpha =  - \rho^{-1}  {\partial \rho(p,T)}/{\partial T}$.
The final equality follows from the cross-derivative of the Gibbs free energy
\cite{TDSM}
\begin{eqnarray}
\frac{\partial h }{\partial p}
& = &
\frac{\partial^2 (\beta G(N,p,T)) }{\partial p\partial \beta}
\nonumber \\ & = &
\frac{\partial (\beta v) }{\partial \beta}
\nonumber \\ & = &
v - \alpha v T .
\end{eqnarray}
Here and throughout $\beta = 1/k_\mathrm{B}T$,
with $k_\mathrm{B}$ being Boltzmann's constant.
With this fountain pressure integral can be translated from the fountain path
to the saturation path,
\begin{eqnarray} \label{Eq:HL-2}
p_B - p_A
& = &
\oint_{T_A}^{T_B} \mathrm{d}T'\;
\rho(T') s(T')
 \\ & \approx &
\int_{T_A}^{T_B} \mathrm{d}T'\;
\Big\{
\rho^\mathrm{sat}(T') s^\mathrm{sat}(T')
\nonumber \\ && \mbox{ }
- \alpha^\mathrm{sat}(T') \big[p(T';p_A,T_A)-p^\mathrm{sat}(T')\big]
\Big\} .\nonumber
\end{eqnarray}
The fountain pressure in the second term in the integrand may be approximated
by that given by the uncorrected expression, Eq.~(\ref{Eq:HL-int}),
rather than iterated to self-consistency.
Numerically using measured data
this form is indistinguishable from the
raw H. London\cite{HLondon39} form on the saturation path,
Eq.~(\ref{Eq:HL-int}), as is now not shown.

A third form for the fountain pressure will be tested
based directly on the equality of chemical potentials
rather than the integral of the fountain pressure.
Writing $\mu_B = \mu_B^\mathrm{sat} +(p_B-p_B^\mathrm{sat}) v_B^\mathrm{sat}$,
we have
\begin{eqnarray} \label{Eq:PA-mu}
p_B - p_A & \approx &
p_B^\mathrm{sat} - p_A
+ \rho_B^\mathrm{sat} (\mu_B - \mu_B^\mathrm{sat})
\nonumber \\ & = &
p_B^\mathrm{sat} - p_A
+ \rho_B^\mathrm{sat} (\mu_A - \mu_B^\mathrm{sat}) .
\end{eqnarray}
This  linear expansion of the exact result must eventually become inaccurate
for large pressure differences.
Invariably the experimental measurements are performed
at saturation of chamber $A$,
and so all quantities on the right hand side,
including $\mu^\mathrm{sat} = h^\mathrm{sat} - T s^\mathrm{sat}$,
can be obtained from standard tables.\cite{Donnelly98}

\begin{figure}[t]
\centerline{ \resizebox{8cm}{!}{ \includegraphics*{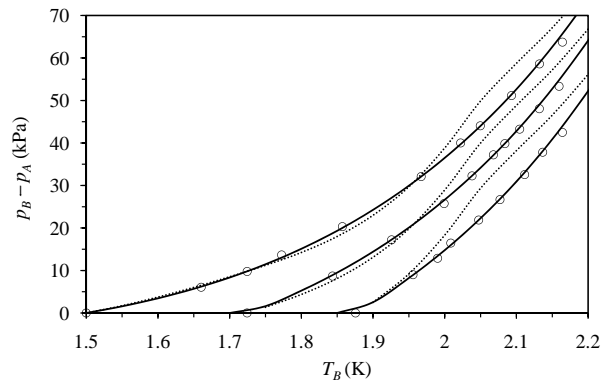} } }
\caption{\label{Fig:fountain}
Measured and calculated fountain pressure for $T_A =$
1.502\,K (left), 1.724\,K  (middle), and 1.875\,K (right).
The symbols are measured data,\cite{Hammel61}
the full curve is the saturation line integral form
of the H. London expression, Eq.~(\ref{Eq:HL-int}),\cite{HLondon39}
and the dotted curve is for fixed chemical potential
using the incompressible fluid
estimate for its departure from the saturation value,  Eq.~(\ref{Eq:PA-mu}).
The calculated curves use measured saturation data
from Ref.~\onlinecite{Donnelly98}.
}
\end{figure}

Figure~\ref{Fig:fountain}
tests the various equations for the fountain pressure
against the measured values.\cite{Hammel61}
The calculations use measured data for $^4$He on the saturation curve,
including the calorimetrically measured entropy.\cite{Donnelly98}
The chemical potential for Eq.~(\ref{Eq:PA-mu})
was obtained from the measured enthalpy and entropy.

It can be seen that the H. London expression
for the temperature derivative of the fountain pressure,
evaluated as an integral along the fountain curve,
Eq.~(\ref{Eq:HL-int}), is practically indistinguishable
from the measured values.
Adding the correction for the translation
from the fountain path to the saturation path,
Eq.~(\ref{Eq:HL-2}),
makes a difference of about $-0.5$\% at the highest fountain pressure shown,
which would be indistinguishable from the uncorrected result,
Eq.~(\ref{Eq:HL-int}), on the scale of the figure.

The equation for the fountain pressure derived
directly from the equality of chemical potential,
Eq.~(\ref{Eq:PA-mu}),
does not perform as well as the integral form
of the H. London\cite{HLondon39} expression,
particularly at higher fountain pressures.
This is in part due to the fact that a linear expansion has been used
to map from the pressure of the second chamber to the saturation pressure.
And in part it is due to measurement error,
which is always magnified by single point applications
and taking the differences between measured quantities.
Integral formulations such as Eq.~(\ref{Eq:HL-int})
tend to smooth out random measurement error.
That the overestimate of the predicted pressure for $T_B \agt 2$\,K
is similar in all three cases despite the differences
in fountain pressures of about a factor of three
tends to suggest that the error is not due to the linearization approximation.
Rather it appears that $\mu_B^\mathrm{sat}$ is systematically too small,
which says that either the measured enthalpy is too small,
or else the measured entropy is too large in this regime.
Despite the limited accuracy of Eq.~(\ref{Eq:PA-mu}),
the agreement with the measured data in Fig.~\ref{Fig:fountain}
is good enough to confirm the thermodynamic analysis that
$\mu_A=\mu_B$ is mathematically equivalent to
$\mathrm{d} p_B/ \mathrm{d} T_B = \rho_B s_B$.

%
\section{Osmotic Pressure in an Incompressible Superfluid}
\label{Sec:Osmo}
\setcounter{equation}{0} \setcounter{subsubsection}{0}
%

In Appendix~\ref{Sec:IdSoln}
the conventional osmotic pressure for a binary solution
is obtained using classical ideal solution theory.
That result may be compared and contrasted with the present results for
a superfluid obtained similarly at the level of an ideal solution.


Consider a system consisting of two chambers at different temperatures
below the condensation temperature and containing $^4$He.
We give the free energy for a single chamber,
and differentiate it to obtain an expression for the chemical potential.
This allows us to equate chemical potentials
and to obtain the fountain pressure.

The constrained Helmholtz free energy for a single system
with $N_0$ ground momentum state bosons and $N_*$ excited momentum state bosons
is given by
\cite{Attard22a,Attard21}
\begin{equation}
-\beta F(N_0| N ,V,T)
=
\ln \frac{\Lambda^{-3N_*} Q}{N_*!V^{N_0}}
+ \sum_{l=2}^\infty  \frac{N_*^l}{N ^{l-1}} g^{(l)} .
\end{equation}
Here $N=N_0+N_*$, $\beta \equiv 1/k_\mathrm{B}T$,
and $\Lambda \equiv \sqrt{2\pi \beta \hbar^2/m}$ is the thermal wavelength.
The series contains the loop grand potentials,
which is a quantum effect that arises from
the symmetrization of the wave function.
The logarithmic term is the classical or monomer term,
and it contains the configuration integral for interacting bosons, $Q(N,V,T)$.
Compared to classical statistical mechanics of a binary mixture
(see Appendix~\ref{Sec:IdSoln}),
for the condensed boson case there is no $N_0!$ in the denominator,
and instead of $\Lambda^{3N_0}$ there appears $V^{N_0}$.
These differences are a consequence
of the non-locality of the permutations of ground momentum state bosons.
\cite{Attard22a,Attard21}

We make the incompressible liquid approximation, $V = Nv$.
The constrained Gibbs free energy, $G = F + pV$,
for chamber $A$ is then given by
\begin{eqnarray}
\lefteqn{
G(N_{0A}|  N_A ,p_A,T_A)
}  \\
& = &
\beta_A^{-1}\left[
N_{*A} \ln \frac{N_{*A} \Lambda_A^3}{N_{A} v_A} - N_{*A}
- \ln \frac{Q_A}{V^{N_A}} \right]
\nonumber \\ && \mbox{ }
- \sum_{l=2}^\infty
\frac{ \beta_A^{-1} N_{*A}^{l}}{N_{A}^{l-1}} g_A^{(l)}
+ p_A N_{A}v_A  .\nonumber
\end{eqnarray}
Recall that $N_A = N_{*A} + N_{0A}$.

Now for an incompressible fluid, since $\mu^\mathrm{ex}=p^\mathrm{ex}v$
(see Appendix~\ref{Sec:IdSoln}),
the derivative of the configuration integral term vanishes,
\begin{eqnarray}
\lefteqn{
\frac{\partial  }{\partial N} \ln \frac{Q(N,V,T)}{V^{N}}
} \nonumber \\
& = &
\frac{\partial (-\beta F^\mathrm{ex}(N,V,T)) }{\partial N}
+ v
\frac{\partial (-\beta F^\mathrm{ex}(N,V,T)) }{\partial V}
\nonumber \\ & = &
-\beta\mu^\mathrm{ex} + \beta p^\mathrm{ex}v = 0.
\end{eqnarray}
This result effectively removes the interactions between the helium-4 atoms,
and makes the analysis more or less the same as that for an ideal gas,
apart from the way in which the fraction of ground and excited
momentum state bosons is calculated or measured.
Consequently, inserting the incompressible fluid assumption at this stage
of the analysis is a rather serious approximation.
It would be better to calculate the chemical potential for interacting atoms,
which is quite feasible,
and this should be explored in the future.

The derivative of the constrained Gibbs free energy
gives the chemical potential,
which with the vanishing of the derivative of the configurational integral,
is
\begin{eqnarray}
\lefteqn{
\mu_{0A} \equiv
\left(\frac{\partial G_A }{\partial N_{0A}} \right)_{N_{*A},p_A,T_A}
}\nonumber  \\
& = &
\frac{- \beta_A^{-1} N_{A*}v}{V_A}
+  p_A v_A
+ \beta_A^{-1} \sum_{l=2}^\infty (l-1)
\frac{  N_{*A}^{l} g_A^{(l)} }{ N_{A}^{l} }
\nonumber \\ & = &
\frac{ - f_{*A}}{\beta_A}
+ p_A   \rho_A^{-1}
+  \beta_A^{-1} \sum_{l=2}^\infty (l-1)  (f_{*A})^l g_A^{(l)} .
\end{eqnarray}
Here $f_{*A} = N_{*A}/N_A$ is the fraction of bosons in excited states
in chamber $A$.
One can similarly obtain an expression for the excited state chemical potential,
\begin{eqnarray}
\mu_{*A}
& = &
\frac{\partial G_A}{\partial N_{*A}}
 \\ & = &
\beta_A^{-1}\ln \frac{N_{*A} \Lambda_A^3}{V_A}
- \beta_A^{-1} \frac{N_{*A} v}{V_A}
+ p_A v_A
\nonumber \\ && \mbox{ }
+  \beta_A^{-1} \sum_{l=2}^\infty
 g_A^{(l)} \left\{(l-1) \frac{N_{*A}^l}{N_A^{l}}
-   l \frac{N_{*A}^{l-1}}{N_A^{l-1}}  \right\}
 .\nonumber
\end{eqnarray}
In the equilibrium state, $\mu_{0A}= \mu_{*A}$,
in which case one must have
\begin{equation} \label{Eq:f*}
\ln  [f_{*A} \rho_A \Lambda_A^3]
=
\sum_{l=2}^\infty  l  g_A^{(l)}  f_{*A}^{l-1}   .
\end{equation}
Since the right hand side is a series of positive terms,
one must have that $f_{*A} > 1/\rho_A \Lambda_A^3$,
which places a lower bound on the fraction of excited state bosons (He~I)
as $T \rightarrow 0$.
This result is predicated
on the incompressible liquid, ideal solution approximation
and also upon the no mixing approximation.

Similar expressions hold for chamber $B$.
Hence equating the chemical potential of the two chambers,
$\mu_{0A}= \mu_{0B}$,
yields the fountain pressure as
\begin{eqnarray} \label{Eq:pAB-osmo}
\lefteqn{
p_A  -  p_B
}  \\
& = &
\rho_A k_\mathrm{B}T_A  f_{*A} - \rho_B k_\mathrm{B}T_B f_{*B}
\nonumber \\ && \mbox{ }
- k_\mathrm{B}\sum_{l=2}^\infty (l-1)
\Big\{ \rho_A  T_A  g_A^{(l)} f_{*A}^l
- \rho_B T_B g_B^{(l)}  f_{*B}^l
\Big\} .\nonumber
\end{eqnarray}
Ignoring the loop terms,
and taking into account the fact
that the excited momentum state fraction increases with increasing temperature,
one sees that the higher temperature chamber has the higher pressure.
Again neglecting the loop terms,
notice the difference between this and the usual osmotic pressure equation
(\ref{Eq:Osmo}).
It is only in  the linear regime that they are the same.

The left hand side can be written as a height difference,
$[ p_A  -  p_B]  \rho^{-1}
\approx
 mg [h_A-h_B]$, where $m$ is the molecular mass.
(Note that here $g$ is the acceleration due to gravity,
not the specific Gibbs free energy,
and $h$ is the height, not the specific enthalpy.)


Extensive calculation of the intensive loop Gaussians $g^{(l)}$
up to $l=7$ have been performed
for a Lennard-Jones model of $^4$He,\cite{Attard21,Attard22a}
and also for $^3$He.\cite{Attard22c}
The fraction of excited state bosons following condensation
has been calculated in the so-called unmixed approximation.
\cite{Attard22a}
These have been used in the present expression for the pressure difference.
The $\lambda$-transition in Lennard-Jones helium-4 occurs at $T=0.65$.
\cite{Attard22a}

\begin{figure}[t]
\centerline{ \resizebox{8cm}{!}{ \includegraphics*{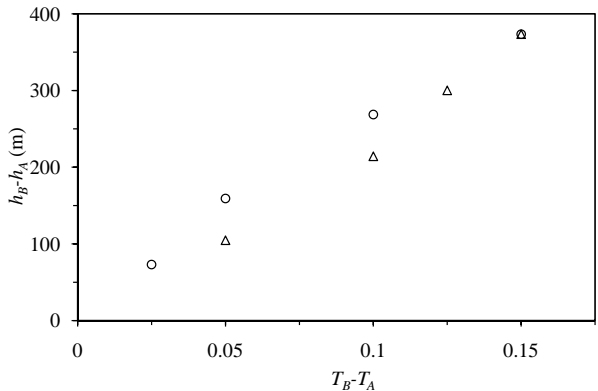} } }
\caption{\label{Fig:height}
The height difference as a function of the temperature difference
for fixed $T_B = 0.65$ (circles) and for fixed $T_A = 0.50$ (triangles)
for Lennard-Jones $^4$He,
for which the temperature in K is $10.22T$.
}
\end{figure}

Figure~\ref{Fig:height} shows the calculated height difference
 as a function of the low temperature $T_A$
for fixed high temperature $T_B = 0.65$,
which is the $\lambda$- temperature in  Lennard-Jones $^4$He.
It also shows the height difference for fixed low temperature $T_A = 0.50$,
It can be seen that the height difference increases
with increasing temperature difference.
The magnitude of the fountain effect is surprisingly large
considering that the largest temperature difference
shown is about 1.5\,K.
The loop series contribution to the total height difference
is on the order of 1--5\%.
This means that the major part of the height differences
in Fig.~\ref{Fig:height} come from the first line
of the right hand side of Eq.~(\ref{Eq:pAB-osmo}),
which are essentially ideal solution terms.

It is noticeable in the figure that for
the same temperature difference, $T_B-T_A = 0.05$,
the higher temperature $T_B=0.65$ has a height difference $h_B-h_A=159.2$\,m,
which is about 50\% larger than that at the lower temperature
$T_A=0.50$, which has a height difference $h_B-h_A=104.7$\,m.
This shows that there is a significant decrease in the
ratio of height difference to temperature difference with decreasing temperature.
This is because the fraction of excited state bosons
decreases with decreasing temperature,
and to leading order the height difference is proportional
to the difference in the excited fraction.

The experimentally measured value
at $T_A=1.143$\,K is 46.8\,m  for $T_B-T_A=0.6$\,K.\cite{Hammel61}
The calculated height difference in Fig.~\ref{Fig:height}
is 104.7\,m for  $T_B-T_A = 0.05 \approx 0.51$\,K,
which is about a factor of 2 too large.
The main reason for the disagreement between the calculated
and the measured height difference
appears to be the use of the incompressible liquid, ideal solution
approximation for the calculations.

\begin{figure}[t]
\centerline{ \resizebox{8cm}{!}{ \includegraphics*{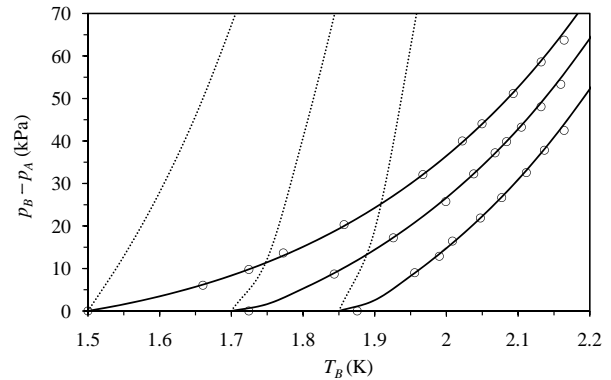} } }
\caption{\label{Fig:fountain2}
Measured and calculated fountain pressure for $T_A =$
1.502\,K (left), 1.724\,K  (middle), and 1.875\,K (right).
The symbols are measured data,\cite{Hammel61}
the full curve is the saturation line integral form
of the H. London expression, Eq.~(\ref{Eq:HL-int}),\cite{HLondon39}
and the dotted curve uses the incompressible fluid
osmotic pressure result,
the first line of Eq.~(\ref{Eq:pAB-osmo}),
using the measured fraction of excited state bosons.\cite{Donnelly98}
The calculated curves use measured saturation data
from Ref.~\onlinecite{Donnelly98}.
}
\end{figure}

Figure~\ref{Fig:fountain2} compares
the incompressible fluid osmotic pressure result,
the first line of Eq.~(\ref{Eq:pAB-osmo}),
with the measured fountain pressure.
This uses the measured\cite{Donnelly98}
 rather than calculated excited state fraction.
It neglects the loop contributions,
which as discussed for the previous figure are expected to be small.
From the fact that the measured curves are nearly  parallel,
one can conclude that the fountain pressure
is approximately a function of the temperature difference,  $T_B -T_A$,
rather than of $T_B$ and $T_A$ individually.
In any case it can be seen that the incompressible,
ideal solution result for the osmotic pressure  performs quite badly.

\begin{figure}[t]
\centerline{ \resizebox{8cm}{!}{ \includegraphics*{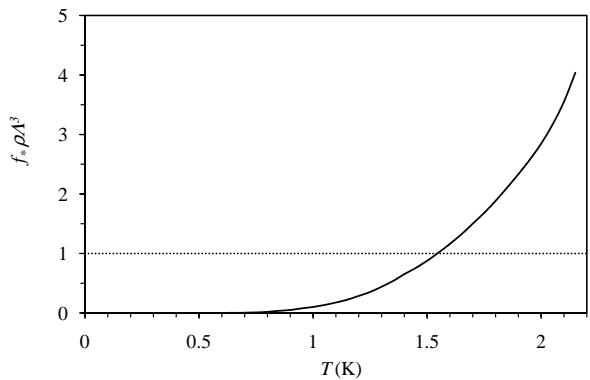} } }
\caption{\label{Fig:f*}
Measured\cite{Donnelly98} fraction of He~I
times the measured\cite{Donnelly98} number density
and thermal wavelength cubed (solid curve).
The dotted line is a guide to the eye.
}
\end{figure}

Figure~\ref{Fig:f*} shows the dimensionless parameter
$f_* \rho \Lambda^3$ using measured data for the fraction of He~I.
\cite{Donnelly98}
As discussed following Eq.~(\ref{Eq:f*}),
equilibrium between ground and excited state bosons
implies that this must be greater than unity,
which the figure shows is violated for $T \le 1.5$\,K.
The tests of the osmotic pressure prediction for the fountain pressure
in Fig.~\ref{Fig:fountain2} are for $T \ge 1.5$\,K.
Nevertheless that $f_* \rho \Lambda^3 < 1$  for $T \le 1.5$\,K
is presumably an indication of the failure
of the incompressible liquid, ideal solution,
and no mixing approximations in the low temperature regime.

\section{Conclusion}

The osmotic pressure mechanism offered by Tisza\cite{Tisza38}
as an explanation for the fountain pressure is problematic on physical grounds.
In the usual realization of osmotic pressure
the solute and the solvent have distinct identities.
Hence the only way to equalize the solvent chemical potential
is to increase the pressure of the high concentration chamber.
In the case of $^4$He,
the ground state bosons can become excited state bosons, and \emph{vice versa}.
Hence if the different fractions in each chamber somehow corresponded to
a chemical potential difference due to mixing entropy,
they could equalize chemical potential
simply by changing their state without changing the pressure.

The present calculations show that the osmotic pressure mechanism
can be an order of magnitude or more too large for the fountain pressure.
Admittedly this is using the incompressible fluid, ideal solution approximation.
But since the common understanding of osmotic pressure
is that it is due to mixing entropy,
and since the latter can be calculated exactly by ideal combinatorics,
the failure of the present ideal solution calculations for the fountain pressure
argue against osmotic pressure as the physical basis of the fountain effect.

It was found here that H. London's\cite{HLondon39} expression
for the temperature derivative of the fountain pressure
corresponds to chemical potential equality of the two chambers.
This follows from the minimization of energy at constant entropy,
which suggests a general principle for superfluid flow
and offers an alternative physical explanation for the fountain effect.

Imagine that the high temperature chamber $B$ is initially at saturation.
For $^4$He,
on the saturation curve the  chemical potential
from the measured\cite{Donnelly98} enthalpy and entropy
decreases with increasing temperature,
$\mathrm{d} \mu^\mathrm{sat} /\mathrm{d} T < 0$.
Hence $\mu_B^\mathrm{sat} < \mu_A^\mathrm{sat}$.
Since $\partial \mu  /\partial p = v > 0$,
the chemical potential of the high temperature chamber can be increased
to achieve equality by increasing the pressure beyond its saturation value.
This occurs as more bosons arrive in the second chamber
because each boson occupies a certain impenetrable volume
(i.e.\ the compressibility is positive).
These are the reasons why condensed bosons initially flow
down the chemical potential gradient from the low temperature chamber
to the high temperature chamber,
and why the high temperature chamber subsequently settles at a higher pressure.

One could describe the new principle
---that the energy at constant entropy
is minimal with respect to particle transfer---
as an empirical formula,
since the final formula agrees quantitatively with measured results
but the axiomatic basis for energy minimization is not well-established.
Undoubtedly one could argue that condensed bosons in the superfluid state
are in a state of low or zero entropy,
and so it would make sense that they respond to mechanical forces
rather than to statistical or entropic forces.
Also the fact that superfluid flow is flow without viscous dissipation
means that the motion of condensed bosons does not produce entropy
and so it cannot respond to entropy gradients.
And of course condensed bosons in a single quantum state
behave as a single particle,\cite{Attard21,Attard22a}
which particle responds as a whole to mechanical forces
without changing the entropy of the system.
Perhaps time and familiarity will raise the status of the present result
from empirical to axiomatic.

It is unlikely that the principle of energy minimization
at constant entropy is restricted to the fountain effect.
Presumably it is a fundamental principle that
applies generally for motion without dissipation in Bose-Einstein condensates.
As such it provides new insight for the understanding and the analysis
of superfluid flow and of superconductor currents.




\appendix

%
\section{Critique of H. London's Derivation} \label{Sec:antiLondon}
\setcounter{equation}{0} \setcounter{subsubsection}{0}
\renewcommand{\theequation}{\Alph{section}.\arabic{equation}}
%

As part of the preparation for this paper,
a detailed study of  H. London's derivation\cite{HLondon39}
of his expression was undertaken,
and the conclusions are reported here. 
The criticisms of the derivation are rather serious
and suggest a disconnect with the final result.
The reader may be sceptical on this point,
given the overwhelming experimental evidence for the quantitative accuracy
of the H. London expression.
And in any case some may wonder whether details of the derivation matter
given that the final expression works.
But it seems to me that the mathematical derivation does matter because
it establishes the status of the final expression:
is it a formally exact thermodynamic result,
or is it only an accurate approximation?
If a theoretical expression is to be used to calibrate instruments,
or as a benchmark against which to judge different experimental techniques,
then it must be provably exact, not merely an approximation
accurate to a limited extent over a limited regime.
Establishing the precise status of the H. London expression
is particularly important in the case of liquid helium
where measured data are reported to four or even more significant figures.
\cite{Donnelly98}

The details of the derivation of H. London\cite{HLondon39}
also matter because
they set a precedence for proper thermodynamic manipulation,
both in general and in the particular case of superfluid flow.
The particular combination of artificial modeling, neglected terms,
and implicit assumptions may well give the correct answer on this occasion,
but not more generally.
Of course since quantitative values of fountain pressure measurements
were known to H. London at the time,\cite{HLondon39}
the fact that his final expression fits the measured data
provides little comfort that his derivation of that expression is sound.
Presumably to some extent he worked backwards,
trying different models and neglecting different terms
until he got a result that fitted the known data.

\subsection{Artificial Model}

\begin{figure}[t]
\centerline{ \resizebox{8cm}{!}{ \includegraphics*{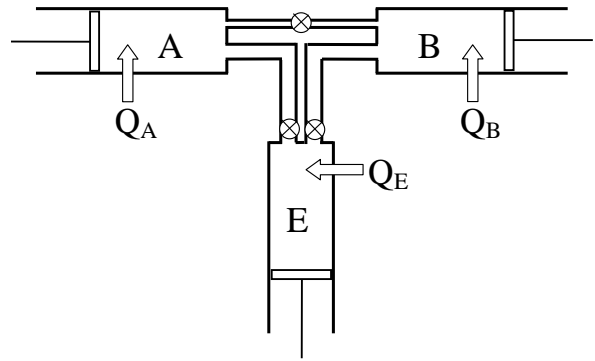} } }
\caption{\label{Fig:model}
Model system analyzed by H. London.\cite{HLondon39}
}
\end{figure}

The model used by H. London\cite{HLondon39}
for his derivation of the fountain pressure expression
is shown in Fig.~\ref{Fig:model}.
In the first half of the cycle,
with the valves to the engine closed,
the pistons on the chambers are used to transfer $N$
bosons from $A$ to $B$ via the capillary.
In the second half of the cycle,
sequentially operating the valves and the pistons,
the $N$ bosons are transferred back from $B$ to $A$ via the engine
following the fountain path $p'=p(T';p_A,T_A)$.
Heat flows in to or out of the helium and work is done on or by it
at each stage.

The first criticism of H. London's derivation\cite{HLondon39}
is the artificiality of this model.
What has this got to do with the actual experimental arrangement
for measuring the fountain pressure?
Are the various heat flows that occur in the artificial engine
the same as those that occur in the actual measurement setup?
What exactly is superfluid about the model?

My second criticism is that
this artificial model seems designed to avoid using
an appropriate thermodynamic potential for the system,
such as the Helmholtz or Gibbs free energy.
These free energies are entirely absent from H. London's paper.\cite{HLondon39}
Also absent is the chemical potential,
which is the usual thermodynamic variable involved in number exchange.

My third criticism is that
the first mention of a specific superfluid property occurs
after Eq.~(10) of the paper by H. London,\cite{HLondon39}
which is interpreted to mean that the superfluid has zero entropy.
This is after the derivation has been completed;
Eq.~(11) is the expression for the fountain pressure derivative,
the present Eq.~(\ref{Eq:HLondon-dp/dT}).
In neither the model itself or the equations used for the derivation
is any restriction or special form due to superfluidity imposed.
In other words, the result of the derivation should apply
equally to an ordinary fluid.
The only way that the correct fountain pressure derivative
for a superfluid can emerge from the derivation
is by some fortuitous error, or implicit assumption, or neglected contribution.

It is true that early in the derivation  H. London\cite{HLondon39}
states that he is treating reversible and irreversible effects
as separable and independent, and that he is neglecting the latter.
But this is not a property confined to superfluids,
since it is a common assumption in irreversible thermodynamics,
and it underpins, for example, the formulation of linear hydrodynamics.
\cite{NETDSM}

\subsection{Thermodynamic and Mathematical Details}

My fourth and final criticism is of the mathematical derivation itself.
There are three key equations:
the work done by the helium,
the change in energy in a cycle,
and the change in entropy in a cycle.

H. London\cite{HLondon39} gives the total work done by the helium in a cycle as
his Eq.~(1a)
\begin{equation}
W = N \oint_{T_A}^{T_B} \mathrm{d}T'\; v' \frac{\mathrm{d}p'}{\mathrm{d}T'}.
\end{equation}
The integral is on the fountain path, $p' = p(T';p_A,T_A)$.
This equation is correct.

H. London\cite{HLondon39} defines the heat flow into the helium
in chamber $A$ from the reservoir as $Q_A$,
and similarly for chamber $B$.
These neglect any energy change via the capillary.
\comment{ 
However one has to distinguish the heat flow into the chamber
from its thermal reservoir, $Q_A'$, and that from the capillary, $Q_A''$,
with $Q_A = Q_A' +Q_A''$, and similarly for chamber $B$.
H. London\cite{HLondon39} effectively sets  $Q_A''=Q_B''=0$.

The capillary heat flow arises from the energy carried off
by the $N$ superfluid bosons,
and the work done by the helium in the chamber in expelling the packet,
$Q_A'' = -N( u_{0A} + p_A v_{0A} ) = - Nh_{0A}$.
In addition to this enthalpic or convective heat flow
that accompanies the transfer of the condensed bosons,
there may also be irreversible conductive heat flow,
which H. London explicitly states he will neglect.
Assuming that no heat flows into the packet during its transit of the capillary,
$Q_B'' = -Q_A''$.
I shall not explore further this point.
} 

H. London\cite{HLondon39} in his Eq.~(2) gives the energy conservation equation,
otherwise known as the First Law of Thermodynamics,
for a full cycle
\begin{equation}
Q_A + Q_B
-
N \oint_{T_A}^{T_B} \mathrm{d}T'\; c^\mathrm{s}_{p_A,T_A}(T')
=
N\oint_{T_A}^{T_B} \mathrm{d}T'\; v' \frac{\mathrm{d}p'}{\mathrm{d}T'}.
\end{equation}
Here $c^\mathrm{s}_{p_A,T_A}(T')
= T' \mathrm{d} s(T';p_A,T_A)/\mathrm{d} T' $
is the heat capacity per particle
along the fountain path (my notation).
Here I've neglected the thermal conductivity in the capillary,
$\mu_\mathrm{cond}=0$, as H. London does later in his analysis.\cite{HLondon39}
The third term on the left hand side is the heat flow into the helium
in the engine.
Since the energy change in the helium in a full cycle is zero,
the heat flow into the helium
must equal the work done by the helium.
This equation is correct.

In actual fact  the integrals can be performed to give
\begin{equation}
Q_A + Q_B + N h_A - N h_B
= 0,
\end{equation}
where $h$ is the enthalpy per particle.
It is a little surprising that H. London\cite{HLondon39}
did not avert to this.

H. London\cite{HLondon39} in his Eq.~(3) gives his version
of the Second Law of Thermodynamics for the cycle,
\begin{equation}
\frac{Q_A}{T_A} + \frac{Q_B}{T_B}
-
N \oint_{T_A}^{T_B} \mathrm{d}T'\; \frac{c^\mathrm{s}_{p_A,T_A}(T')}{T'}
=
0.
\end{equation}
In referring to this equation as the Second Law of Thermodynamics,
H. London\cite{HLondon39} does not appear to be making any statement
about entropy maximization.
Rather he seems to be making reference to the fact
that the heat flow divided by temperature gives a change in entropy.

H. London\cite{HLondon39} appears to believe that this is an equation
for the change in entropy of the subsystems (the chambers) alone,
which by definition must be zero for a full cycle.
But this equation is not the true equation for the change in entropy
of the subsystems as it neglects the work done.
Obviously any $pV$-work on a system
changes its energy and therefore its entropy;
indeed at constant number, $\mathrm{d}S = \mathrm{d}E/T + p\mathrm{d}V/T$.
The contribution of the work done by the helium in the chambers
in changing their volume
is neglected in H. London's\cite{HLondon39} Eq.~(3).

Also there is no conservation law for entropy.
It is one thing to say that the conserved variables such as energy
return to the same state after a full cycle,
and that therefore flows in them have to add to zero;
it is quite another thing to attempt to analyze flows
in non-conserved variables such as entropy.
I don't see the point in attempting to analyze the change in entropy
in a cycle unless one intends to maximize it,
in which case it should be the total entropy of the chambers
and the thermal reservoirs, not just of the chambers.

I also point out the very specific model used by H. London\cite{HLondon39}
for performing work on the helium in the chambers.
Such mechanical pistons do not change the entropy of the reservoirs external
to the chambers.
If one used a more realistic model for the external work,
such as the Gibs free energy to represent fluid external to that
part of the chamber that is the focus,
then one would have to take into account the change in external entropy
as work was done on the helium in the chamber.
In this case one does not arrive at H. London's\cite{HLondon39} result
for the temperature derivative of the fountain pressure.

In summary,
whilst I do not doubt the validity
of the expression for the temperature derivative of the fountain pressure
given by H. London,\cite{HLondon39}
the present Eq.~(\ref{Eq:HLondon-dp/dT}),
I have grave misgivings about the derivation he offers for it.
A theory has to be convincing on its own mathematical and axiomatic merits.
A theory cannot be justified by agreement with experiment
when the process of developing the end product
consists of  tinkering with the derivation  and model
until a fitting result is obtained.
Is the successful `best fit' selected for publication
any more convincing than those models that didn't make it?

In the case of H. London's formula,\cite{HLondon39}
I find his derivation flawed on several grounds.
It appears to me that the combination of the artificial cycle,
the unrealistic mechanical external work,
the accident that the incorrect expression
for the change in entropy of the chambers
actually equals the correct expression for the change
in the thermal reservoirs,
all conspire to give the known experimental result for the fountain pressure
even though no superfluid property is actually input into the derivation.
Because of these flaws in the derivation,
the exact status of the H. London\cite{HLondon39} formula
for the fountain pressure is in my opinion unproven.

%
\section{Incompressible Classical Ideal Gas} \label{Sec:IdSoln}
\setcounter{equation}{0} \setcounter{subsubsection}{0}
\renewcommand{\theequation}{\Alph{section}.\arabic{equation}}
%


Consider two chambers, $A$ and $B$,
connected by a semi-permeable membrane transparent to the solvent 1.
Assume that the solvent and solute are incompressible,
so that the the volume of the first chamber is $V_A = N_{A1} v_1 + N_{A2} v_2$,
and similarly for the second chamber.
The total number is fixed, $N_n = N_{An}+N_{Bn}$, $n=1,2$.
For the classical ideal gas,
the constrained Gibbs free energy is given by\cite{TDSM}
\begin{eqnarray}
\lefteqn{
\beta G(N_{A1}|\underline N ,p_A,p_B,T)
}  \\
& = &
\sum_{n=1}^2 \Big\{
 N_{An} \ln \frac{N_{An} \Lambda^3}{N_{A1} v_1 + N_{A2} v_2} - N_{An}
 \nonumber \\ && \mbox{ }
 +
 N_{Bn} \ln \frac{N_{Bn} \Lambda^3}{N_{B1} v_1 + N_{B2} v_2} - N_{Bn}
 \Big\}
  \nonumber \\ && \mbox{ }
+ \beta p_A[N_{A1} v_1 + N_{A2} v_2] + \beta p_B [N_{B1} v_1 + N_{B2} v_2] .
 \nonumber
\end{eqnarray}
Here $\beta = 1/k_\mathrm{B}T$ is the reciprocal temperature,
$\Lambda = \sqrt{2\pi \hbar^2 \beta/m}$ is the thermal wavelength,
$m$ is the mass, and $\mu$ is the chemical potential.
Since $G = \mu N$,
this shows that for an incompressible fluid,
$\mu^\mathrm{ex}=p^\mathrm{ex}v$.
The derivative is
\begin{eqnarray}
\left(\frac{\partial \beta G }{\partial N_{A1}} \right)_{N_1}
& = &
\ln \frac{N_{A1} \Lambda^3}{N_{A1} v_1 + N_{A2} v_2}
-
\ln \frac{N_{B1} \Lambda^3}{N_{B1} v_1 + N_{B2} v_2}
 \nonumber \\ && \mbox{ }
- \frac{N_{A1}v_1}{N_{A1} v_1 + N_{A2} v_2}
+ \frac{N_{B1} v_1}{N_{B1} v_1 + N_{B2} v_2}
 \nonumber \\ && \mbox{ }
- \frac{N_{A2}v_1}{ N_{A1} v_1 + N_{A2} v_2 }
+ \frac{N_{B2}v_1}{N_{B1} v_1 + N_{B2} v_2}
  \nonumber \\ && \mbox{ }
+ \beta p_A  v_1  -  p_B  \beta v_1
\nonumber \\ & = &
\ln \frac{\rho_{A1}}{\rho_{B1}}
-[\rho_{A1} + \rho_{A2} - \rho_{B1}  - \rho_{B2}] v_1
  \nonumber \\ && \mbox{ }
+ \beta [ p_A  - p_B]  v_1 .
\end{eqnarray}
Hence at equilibrium
\begin{equation}
\ln \frac{\overline \rho_{A1}}{\overline \rho_{B1}}
-[\overline\rho_{A1} + \overline\rho_{A2}
- \overline\rho_{B1}  - \overline\rho_{B2}] v_1
 v_1
=
-\beta [ p_A  - p_B]  v_1 .
\end{equation}
For simplicity
choose $v_1=v_2=v$, so that $\rho = \rho_1+\rho_2 = v^{-1}$
and $\rho_A=\rho_B$.
In this case the equilibrium condition becomes
\begin{equation} \label{Eq:Osmo}
\ln \frac{\overline \rho_{A1}}{\overline \rho_{B1}}
=
\frac{-\beta}{\rho} [ p_A  - p_B]  .
\end{equation}
Hence
$ p_A  \ge  p_B$ implies that $\overline \rho_{A1} \le \overline \rho_{B1}$.
The chamber with the greater pressure is the one with the lower solvent fraction
(and hence higher solute fraction).
This is the classical osmotic pressure.

%
\section{Irreversible Heat Flow} \label{Sec:Irrev}
\setcounter{equation}{0} \setcounter{subsubsection}{0}
\renewcommand{\theequation}{\Alph{section}.\arabic{equation}}
%

The body of the paper has been concerned with the reversible
contributions to the fountain effect,
which determine the structure and pressures of the two chambers.
It is also of interest to explore at an elementary level of approximation
the irreversible contributions, which determine the heat flow
in the steady state.

Assuming the enthalpy of the packet is constant,
then for $N$ bosons making the passage $A\rightarrow B$
via the capillary the heat flow into the helium is
\begin{eqnarray}
Q_B^\mathrm{irrev}
& \approx &
Nu_{B}-[Nu_{A} - (p_B-p_A) N v].
\end{eqnarray}
This assumes that changes in the volume per boson can be neglected.
Assuming that the initial energy per particle of the packet
is representative of the chamber as a whole,
$u_{0A} \approx u_A$,
allows the irreversible heat flow in the starting chamber to be
taken to be zero, $ Q_A^\mathrm{irrev} = 0$.
Obviously these approximations could be refined.

For the return passage $B\rightarrow A$ via the capillary,
the irreversible heat flow is
\begin{eqnarray}
Q_A^\mathrm{irrev}
& \approx &
Nu_{A}-[Nu_{B} + (p_B-p_A) N v] .
\end{eqnarray}
Again the starting energy is taken to be representative,
$u_{0B} \approx u_B$, and
the irreversible heat flow in the starting chamber is set to zero,
$ Q_B^\mathrm{irrev} = 0$.

Now the change in entropy of the universe
upon cycling a packet of $N$ bosons between the chambers is
\begin{eqnarray}
\frac{\Delta^\mathrm{irrev}S_\mathrm{tot}}{N}
& = &
\frac{-Q_B^\mathrm{irrev}}{NT_B}
+
\frac{-Q_A^\mathrm{irrev}}{NT_A}
 \\ & \approx &
[u_B-u_A + (p_B-p_A) v] \frac{T_B-T_A}{T_AT_B}  .\nonumber
\end{eqnarray}
Since both the energy and the pressure increase with temperature,
this is positive and quadratic in the temperature difference, as it must be.
In the steady state,
this drives simultaneous forward and backward superfluid flow
in the capillary.

\newpage

\end{document}